\documentclass[twocolumn,twoside,slac_two]{revtex4}
\usepackage{graphicx}
\usepackage{fancyhdr}
\begin{document}

\title{Prospects of Non-Riemannian Cosmology}

\author{Dirk Puetzfeld}
\affiliation{Department of Physics and Astronomy, Iowa State University, Ames, IA 50011, USA}
\email{dpuetz@iastate.edu}

\begin{abstract}
In this work we provide the motivation for considering non-Riemannian models in cosmology. Non-Riemannian extensions of general relativity theory have been studied for a long time. In such theories the spacetime continuum is no longer described by the metric alone but endowed with additional geometric quantities. These new quantities can be coupled to the intrinsic properties of matter in a very natural way and therefore provide a richer gravitational theory, which might be necessary in view of the recent cosmological evidence for dark matter and dark energy. In this work we mainly focus on the concepts in metric-affine gravity and point out their possible significance in the process of cosmological model building.
\end{abstract}

\maketitle

\thispagestyle{fancy}

\section{Introduction}

Cosmology, especially its observational sector, is currently a thriving field of physics. On the theoretical side opinions have converged to what is nowadays dubbed {\it cosmological concordance model} (CCM). But, despite of all the successes of this model in describing different cosmological observations, we should not fool ourselves to believe that the grand picture of cosmology stands on a firm basis. The reason for this is simple: Interpretation of the data within the concordance model leads inevitably to the introduction of the concepts of {\it dark matter} and {\it dark energy}\footnote{See \cite{Fukugita2004} for an inventory of cosmological parameters.}. We surely could live with such concepts by stating that they depend on some peculiar details which yet have to be added to the description of our universe. Unfortunately, dark matter and energy make up the complete energy budget within our simple picture and therefore cannot be treated as some minor details which remains to be worked out. This is clearly an embarrassing situation which needs to be addressed by cosmologists. 

In the following we will have a glance at the theoretical landscape of cosmology and pay special attention to the non-Riemannian approach. Non-Riemannian extensions of our current gravity theory, i.e.\ General Relativity (GR), represent a well motivated framework and have been discussed extensively in the literature. In this short review we only address questions which are related to the possible cosmological significance of such an approach. Readers who want to learn more about the fundamentals of non-Riemannian gravity theories, the gauge theoretical approach to gravity, and metric-affine gravity (MAG) should consult the excellent reviews \cite{Blagojevic,Goenner2004,Hammond2002,Hehl1976,PhysRep}.

\section{Theoretical landscape}

With the right amount of crudeness one could summarize the reasons to consider a drastic step, like the change of the gravity theory which underlies cosmology, as follows:

\begin{itemize}

\item Large amounts of dark matter/energy necessary to fit current observations within the CCM.

\item No direct observation of a dark matter particle in the laboratory.

\item No theoretical explanation for the smallness of the dark energy component when compared to quantum field theory.

\item No reason to believe that GR is valid in the early universe, i.e.\ at high energies.

\item No test of Newtonian/general relativistic gravity on cosmological scales.

\end{itemize}

In the following we have a glance at some of the proposed remedies for this situation.  

\subsection{Alternatives}

Although we know for sure that GR has to be modified in order to make it compatible with quantum theory \cite{Kiefer}, we do not have any final form of this new gravitational theory. Additionally, we do not know how possible low-energy modifications, and thus modifications that may play an important role for the aforementioned observational problems in cosmology, caused by such a new theory will look like.  

In table \ref{table_models} we provided a very rough overview over current theoretical approaches to extend/replace our current gravity theory and thereby also our cosmological model. The separation into different model classes is sometimes not unique. For example, one could also count the non-symmetric gravity models as non-Riemannian models and all of the models listed in table \ref{table_models} could in principle also have a non-trivial topology.

\begin{table}[t]
\begin{center}
\caption{Some examples for different classes of models recently used to explain observations of cosmological significance.}
\begin{tabular}{|p{3.5cm}|p{4.5cm}|}
\hline \textbf{Model type} & \textbf{Description} \\
\hline Scalar-tensor theories& Modified Lagrangian, additional scalar field (maybe a leftover from some higher theory) non-minimally coupled to the Ricci scalar.\\
\hline Higher dimensions& Our universe represents only a 4-d brane in a 5-d bulk, gravity assumed to be the only interaction which propagates in the bulk.\\
\hline $f(R, R_{\alpha \beta}, \dots)$ models& Modified gravitational Lagrangian in terms of the curvature.\\
\hline Topological models& Universe assumed to have non-trivial topology, i.e.\ impose some additional global properties of spacetime which GR, as a local theory, makes no statements about.\\
\hline Non-symmetric gravity& Theories in which the metric $g_{\alpha \beta}$ is no longer symmetric.\\
\hline Tensor-vector-scalar theory&Additional vector field introduced by hand into the definition of the metric, extended Lagrangian which contains the additional vectorial quantity and an extra scalar field. \\
\hline Non-Riemannian models& Spacetime no longer Riemannian, new field strengths torsion $T^\alpha{}_{\beta \gamma}$ and nonmetricity $Q_{\alpha \beta \gamma}$ couple to intrinsic properties of matter such as the spin.\\
\hline 
\end{tabular}
\label{table_models}
\end{center}
\end{table}

\begin{figure}
\includegraphics[width=80mm]{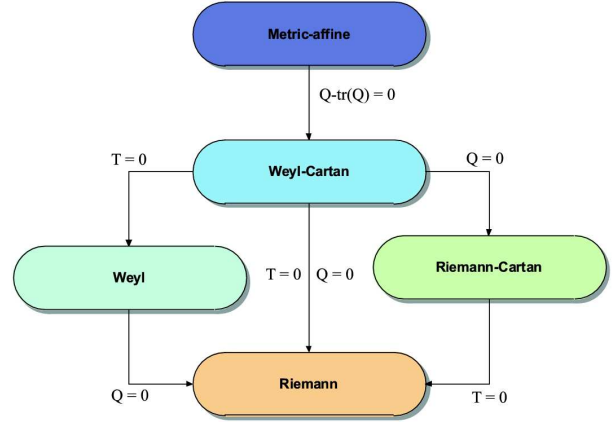}
\caption{Classification of different spacetime types according to the non-Riemannian scheme. By switching off torsion and nonmetricity we arrive at the usual Riemannian spacetime as encountered in GR.}
\label{fig_spacetimes}
\end{figure}

\subsection{Non-Riemannian gravity}

One of the general frameworks for non-Riemannian gravity theories is metric-affine gravity (MAG) as reviewed by Hehl et al.\ in \cite{PhysRep}. In the following we focus on the new geometrical notions in metric-affine gravity and try to explain their possible impact on cosmology. The reasons to pick MAG as starting point are varied: (i) Among the different alternative gravity theories MAG represents a very well motivated and natural generalization, c.f.\ the introduction of \cite{PhysRep} for a list of arguments, and (ii) there exists a general Lagrangian formulation of MAG according to which many other non-Riemannian theories may be systematically classified. (iii) There exist several exact (also non-cosmological) solutions for MAG which rank it among the best studied alternative gravity theories during the last years. (iv) The idea to couple intrinsic features of matter to new geometrical quantities can be viewed as the natural prolongation of the line of reasoning which led to the formulation of the so far most successful gravity theory, namely GR. 

\subsection{Metric-affine gravity}

In metric-affine gravity the spacetime continuum which contains matter carries both stresses $\sigma^{\alpha \beta}$ (or momentum currents) and hyperstresses $\Delta^{\alpha \beta \gamma}$ (or hypermomentum currents). The geometry of spacetime is described by means of a metric $g_{\alpha \beta}$ and an independent affine connection $\Gamma^\gamma_{\alpha \beta}$. The metric is still symmetric, i.e.\ $g_{\alpha \beta} = g_{\beta \alpha}$, but the connection is no-longer given by the metric compatible connection $\left\{ {}^\alpha_{\beta \gamma}\right\}$\footnote{$\left\{ {}^\alpha_{\beta \gamma}\right\}:=\frac{1}{2}g^{\alpha \mu} \left( \partial_\beta g_{\gamma \mu} + \partial_\gamma g_{\beta \mu} -\partial_\mu g_{\beta \gamma} \right)$.} known from GR and may be asymmetric $\Gamma^\gamma_{\alpha \beta} \neq \Gamma^\gamma_{\beta \alpha}$. If we define Cartan's torsion tensor $T^\alpha{}_{\beta \gamma}:=\Gamma^\alpha_{[\beta \gamma]}$ and the nonmetricity tensor $Q_{\alpha \beta \gamma}:=-\nabla_\alpha g_{\beta \gamma}$ then, cf.\ \cite{Schouten}, the affine connection might be split up as follows
\begin{eqnarray}
\Gamma^\alpha_{\beta \gamma}=\left\{ {}^\alpha_{\beta \gamma}\right\} &+& T_{\beta \gamma}{}^{\alpha} - T_\gamma{}^\alpha{}_\beta + T^\alpha{}_{\beta \gamma} \nonumber \\
&+&\frac{1}{2} \left( Q_{\beta \gamma}{}^\alpha + Q_\gamma{}^\alpha{}_\beta - Q^\alpha{}_{\beta \gamma}\right).\label{gencon}
\end{eqnarray}
Furthermore, one assumes that the momentum current $\Sigma^{\alpha \beta}$ of matter essentially couples to the metric $g_{\alpha \beta}$ whereas the hypermomentum current $\Delta^{\alpha \beta \gamma}$ couples to the affine connection $\Gamma^\alpha_{\beta \gamma}$ of the spacetime.

From the last assumption and the splitting of the connection as given in eq.\ (\ref{gencon}) it becomes clear that MAG incorporates several other alternative gravitational theories, as well as GR itself. For example it is well known from Einstein-Cartan (EC) theory that the torsion of spacetime couples to the intrinsic spin of particles.  In figure \ref{fig_spacetimes} we sketched how different spacetimes, and thereby different alternative theories which make use of these richer spacetime concepts, maybe classified with respect to the torsion and nonmetricity.

The gravitational field Lagrangian of MAG is expected to be of the form $L=L\left(g,\partial g, \Gamma, \partial \Gamma \right)$ and a matter Lagrangian minimally coupled to the new geometrical fields $L_{\rm m}=L_{\rm m}\left( \psi, \nabla \psi, g\right)$. The gravitational field equations are given by the variational derivatives with respect to the metric $\delta L / \delta g_{\alpha \beta} \sim \sigma^{\alpha \beta}$ and the connection $\delta L / \delta \Gamma^\alpha_{\beta \gamma} \sim \Delta_{\alpha}{}^{\beta \gamma} $. We only mention here that a very general suggestion for the dynamics of this theory, which makes use of a slightly different but equivalent notation than the one used here, has been made in \cite{Exact2}. 

\section{Cosmology}

In \cite{Puetzfeld2004} we provided a brief chronological guide to the literature on non-Riemannian cosmological models. Therein the developments in cosmology were traced back to the early seventies and were given in table form. Most of the early non-Riemannian cosmological models were based on Einstein-Cartan theory. Investigations mainly revolved around the construction of exact solutions and the question of whether or not an initial singularity can be avoided in such models. In the 1980s more general types of Lagrangians were considered. The inclusion of quadratic terms in the Lagrangian, leading to dynamical degrees of freedom, was mainly motivated by the framework of Poincar\'{e} gauge theory (PGT) and led to new classes of exact solutions. The story continued with the advent of the inflationary model, which led to investigations which tried to mimic or justify this new idea within different non-Riemannian scenarios. Till the end of the 1990s most of the works in non-Riemannian cosmology (NRC) were focused on the description of the early stages of the universe. This bias can mainly be ascribed to the estimates for the new spin-spin contact interaction encountered in Einstein-Cartan theory. This interaction shows up at extremely high energy densities\footnote{In \cite{Hehl1973,Hehl1976} it was estimated that this may be the case at approximately $10^{47}\, {\rm g/cm}^3$.} and might therefore play only a crucial role in the early universe. This focus has changed during recent years, mostly due to the persisting need for the large amount of dark matter and more recently also dark energy. The requirement of dark energy at late stages of the cosmic evolution might be taken as an indicator for the presence of new physics possibly due to some non-Riemannian relics in cosmology.

Since the field equations of the FLRW model are extremely simple and the main evidence (see \cite{Barris, Tonry} for the latest SNIa samples) for the new dark energy component comes from cosmological tests which are related to the expansion history of the universe, it is natural to study the impact of changes in this history and their possible origin. There have been several suggestions for modifications during the last years coming from different directions, cf.\ table \ref{table_models}. Non-Riemannian models, especially MAG, provide a very good starting point for the study of changes of the cosmological field equations which are justified by a grander theory, in the case of MAG by the general Lagrangian provided in \cite{Exact2}. Let us now come to the rhs, i.e.\ the matter side, of the field equations.

\subsection{Continua with microstructure}

\begin{figure}
\includegraphics[width=80mm]{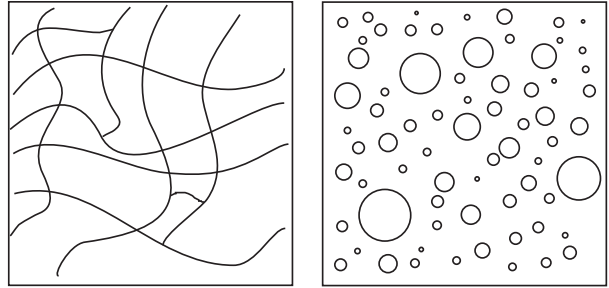}
\caption{In more complex fluid models matter may be represented by a medium with dislocations or finely dispersed voids. Non-Riemannian models allow for a natural coupling of the new geometrical quantities like torsion and nonmetricity to such kind of fluid properties.}
\label{fig_dislocbub}
\end{figure}

\begin{figure*}[t]
\centering
\includegraphics[width=100mm]{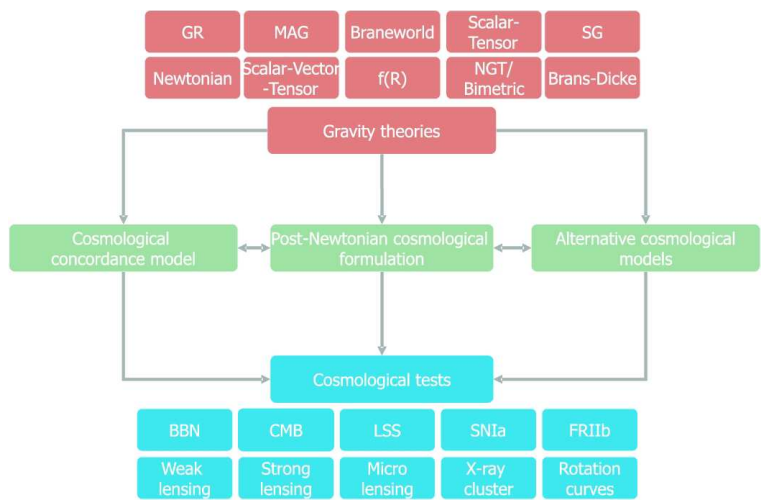}
\caption{How to build and compare cosmological models?} \label{fig_modelbuild}
\end{figure*}

A very promising approach to construct cosmological models in a non-Riemannian setup is related to the availability of more sophisticated fluid models. In the case of metric-affine gravity Obukhov \& Tresguerres \cite{Obukhov1993,Obukhov1996} devised a fluid model termed the {\it hyperfluid} \footnote{See also \cite{Babourova1998}.}. This kind of fluid can be used as a natural source for the hypermomentum current which appears on the rhs of the field equations of MAG. The new degrees of freedom in such a fluid model can be coupled to the new geometrical properties, i.e. torsion and nonmetricity. In the hyperfluid picture, which can viewed as a generalization of early spin-fluid models \cite{Weyssenhoff,Kopczynski,Obukhov1987,deRitis1983,deRitis1985,Ray}, the motion of the fluid is described by usual four-velocity and a triad attached to each fluid element, which can undergo arbitrary deformations during the motion of the fluid. This is analogous to the description of continua with microstructure \cite{Capriz} in the theory of elasticity. In figure \ref{fig_dislocbub} we sketched two examples of such media, namely one with dislocations and another one with finely dispersed voids. The hyperfluid and special cases of it were used in several cosmological models in the past\footnote{See \cite{Puetzfeld2004} for a list of references.}, a systematic treatment for a fairly general Lagrangian of MAG will be published in \cite{Puetzfeld2005}.

\section{How to test and compare?}

As we sketched in figure \ref{fig_modelbuild}, the list of different cosmological tests is quite long, and still growing. Usually one starts with a theoretical model from the upper portion of the figure and then compares it to some of the observations in the lower portion of the figure in order to falsify it. In view of the sometimes very different theoretical approaches this can become quite cumbersome, i.e.\ one has to spend a lot of time to work out the single tests in a scenario which significantly deviates from the cosmological concordance model. Therefore one of the most pressing tasks in cosmology is the development of a post-Newtonian framework which allows us to compare different theoretical approaches in a systematic and somewhat standardized way, and hopefully allows for the a fast backreaction of the cosmological tests on the theoretical model. In figure \ref{fig_modelbuild} we denoted such a framework, which has yet to be developed, by the connecting middle part.

\section{Conclusion \& Outlook}

Up to now there seems to be no real competitor model in the non-Riemannian context which can replace the current cosmological concordance model and at the same time explain the effects caused by dark matter/energy in a purely gravitational way. Most of the models proposed so far are either not worked out to a sufficient level of detail, fail one or more of the cosmological tests, or are not distinguishable from the CCM with the currently available data. But, and this cannot be stressed enough, we are clearly only beginning to explore the different possibilities of non-Riemannian models. This is especially true for the cosmological sector of metric-affine gravity which currently only covers a very small region of the theoretically permissible parameter space.  

\bigskip

\begin{acknowledgments}
The author wants to thank F.W.\ Hehl for constant advice and support. The financial support by M.\ Pohl and SLAC is gratefully acknowledged. Additionally, the author wants to thank S.\ LeBohec for the hospitality during his stay at University of Utah.
\end{acknowledgments}

\end{document}